\documentclass[prl,twocolumn,preprintnumbers,amsmath,amssymb]{revtex4}

\usepackage[dvips]{graphicx}
\usepackage{dcolumn}
\usepackage{bm}

\begin{document}

\preprint{UAB--FT--582}

\title{Popper's test of quantum mechanics and two-photon ``ghost'' diffraction}

\author{Albert Bramon}
\email{bramon@ifae.es}
\affiliation{Grup de F\'{\i}sica Te\`orica, Universitat Aut\`onoma de Barcelona, E-08193 Bellaterra (Barcelona), Spain}

\author{Rafel Escribano}
\email{Rafel.Escribano@ifae.es}
\affiliation{Grup de F\'{\i}sica Te\`orica and IFAE, Universitat Aut\`onoma de Barcelona, E-08193 Bellaterra (Barcelona), Spain}


\begin{abstract}
A test on quantum mechanics proposed long ago by Karl Popper is reconsidered
with further detail and new insight.
An ambiguity in the proposal, which turns out to be essential in order to make the test conclusive,
is identified and taken into account for the first time.
Its implications for recently performed photon experiments
[such as in D.~V.~Strekalov \textit{et al.}, Phys.\ Rev.\ Lett.\ {\bf 74}, 3600 (1995)]
are briefly analyzed. 
\end{abstract}

\maketitle

Quantum Mechanics (QM) plays an essential role in our present understanding
of many fundamental physical phenomena.
Most of them manifest highly counterintuitive behaviours and the corresponding
QM explanation is full of subtle and unfamiliar concepts. 
As a consequence, misconceptions and controversies have been 
---and continue to be---
quite long and frequent.
An excellent example is the thought experiment that Karl Popper proposed and
continued to improve between 1934 and 1987  \cite{Popper34,Popper82,Popper84,Popper87}.
In the most recent versions, Popper makes a twofold assertion: 
\textit{i)} that his is a \textit{crucial} test  \cite{Popper82,Popper87} on the interpretation
of QM and \textit{ii)} that he ``is inclined to predict'' that, if experimentally performed,
the test would decide against the (subjectivist) Copenhagen interpretation \cite{Popper82}.
Needless to say, such statements have raised a considerable and interesting controversy
\cite{Krips,BS,CL,Peres,Plaga,Short,Qureshi1,Qureshi2,Qureshi3},
both on the conclusive character of the test and on its relation with recently performed experiments
on two-photon ``ghost'' diffraction \cite{Kim:1999td,Strekalov}. 

The purpose of our paper is to contribute to clarify these issues:  
\textit{i)} by identifying an ambiguity in Popper's proposal which,
in spite of being essential in order to obtain the correct QM predictions,
it has not been explicitly considered in previous analyses and 
\textit{ii)} by establishing the connection between Popper's test and the two-photon
diffraction experiment by Strekalov \textit{et al.}~\cite{Strekalov}. 

The experimental setup proposed by Popper is quite simple.
Its projection on the plane comprising the horizontal $x$-axis and the vertical $y$-axis
is schematically represented in Fig.~1.
Two vertical and parallel screens are placed on the planes $x = \pm d$
left and right from the origin, Fig.~1a. 
There is a slit on each screen which allows the passage of
photons with vertical coordinates in a narrow range, $-a \le y \le +a$.
Photons are emitted in pairs from a source $S$ placed at the origin 
---positronium decays into two $\gamma$'s, in Popper's original proposal---
 and the analysis restricts to photon pairs whose members are both jointly detected. 
Some  pairs will indeed have both members passing through the slits and  the
vertical momenta of these left- and right- moving photons,
$(k_1 )_{y} \equiv k_{1}$ and $(k_2 )_{y} \equiv k_{2}$,
will finally be measured by two arrays of detectors placed at distances $D$ far away
behind the slits, $D\gg a$. 
For suitable slit widths and photon momenta, 
$|\vec{k}_{1,2}| \sim \pi / a$, 
single-slit diffraction theory predicts $\Delta k_{1,2}\simeq 1/2a$ for the dispersions
of the vertical components of these momenta. 

Popper further assumes that the source decays at rest and the two photons
are thus emitted with opposite momenta, $\vec{k_{1}} +\vec{k_{2}} =0$.
Thanks to this momentum entanglement between the two members of each pair,
one can analyze what happens in a second run of the experiment when the right-side slit
is wide open or, in practice, the right-screen removed, Fig.~1b.
Popper concludes that, for coincident detections, standard QM predicts that 
the right-side momentum dispersion, $\Delta k_{2}\simeq 1/2a$, 
remains  and can be further increased by narrowing the width $2a$ of the still
\textit{in situ} left-slit.
This prediction follows from the assumption that the passage of the left-moving photon
through the left-slit amounts to performing a measurement of the $y$-coordinate with
$\Delta y_1\simeq a$ and that,
as a consequence of the entanglement and common origin of the two photons,
the vertical position of its  right-moving partner is known with the same accuracy 
$\Delta y_2\simeq a$.
Popper is quite explicit on this point \cite{Popper82}:
\begin{quotation}
To sum up: 
if the Copenhagen interpretation is correct, then any increase in the precision of our
\textit{mere knowledge} of the position $[ y_2 ]$ of the particles going to the right
should increase their scatter; and this prediction should be testable.
\end{quotation}
In other words, for coincident detections a ``ghost'' image of the left-slit performs a kind of 
``indirect measurement'' of $y_2$ and produces ``virtual'' diffraction on the right-side.
The momentum dispersion is predicted to be $\Delta k_{2}\simeq 1/2a$ by considering that,
according to standard QM, Heisenberg's principle is 
``applicable to all kinds of indirect measurements'' \cite{Popper82}. 
By contrast, Popper's objective interpretation of QM 
---based on propensities rather than on subjective knowledge---
predicts a decrease of $\Delta k_{2}$ when the left-side slit is narrowed.
This makes the test a \textit{crucial} one \cite{Popper82}. 

A serious criticism to the preceding analysis has been pointed out by several authors 
\cite{BS,CL,Peres,Plaga,Short,Qureshi1,Qureshi2,Qureshi3}. 
The source $S$ cannot be 
exactly localized at the origin and perfectly at rest, as required to argue that,
for jointly detected events, one  has $\Delta y_1=\Delta y_2\simeq a$ 
regardless the presence (Fig.~1a) or not (Fig.~1b) of the right-slit.  
The undecayed source itself or, equivalently, the global two-photon final system
has to obey Heisenberg's principle and, accordingly, the vertical components of the
CM-position, $(y_1+y_2)/2$, and total momentum, $k_1+k_2$,
must satisfy $\Delta (k_1 + k_2)\Delta\left[ (y_1 + y_2)/2\right]\ge 1/2$.
Once this constraint is imposed, the analyses  of Refs.~~\cite{BS,CL} 
---based on simple geometrical arguments---
claim that Popper's proposal is no longer able to discriminate between the two approaches. 
Similarly, the discussion in \cite{Qureshi1,Qureshi2}
---based on a two-particle state compatible with Heisenberg's principle 
and a simplified assumption on wave-function reduction---
claims that in the case of Fig.~1b standard QM  predicts no increase of $\Delta k_{2}$
when narrowing the left-side slit.
A claim which contradicts Popper's original analysis \cite{Popper82,Popper87}
and would make his test inconclusive as well. 

\begin{figure}
\includegraphics[width=0.45\textwidth]{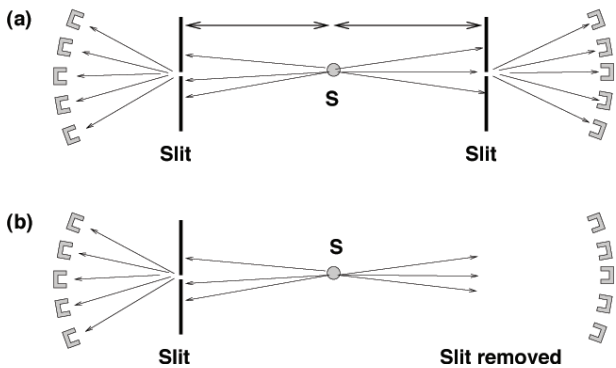}
\caption{\label{fig1}
Schematic diagram of Popper's proposal. 
(a) Two symmetrical narrow slits produce similar single-slit diffraction patterns
for two-photon events detected in coincidence by the left- and right-arrays of detectors.
(b) The slit on the right is wide open (the screen has been removed) 
and the right-side diffraction pattern is analyzed for coincident events as in (a).}
\end{figure}

In spite of all these claims, we have recently argued \cite{Bramon:2005pz} that the above criticisms 
---which are fully reasonable, at least in principle---
do not necessarily destroy the whole ``falsifiability'' potential of Popper's proposal.
The situation is more subtle and deserves further analysis \cite{Peres}.
There is an ambiguity in Popper's proposal 
---a new feature which has not been considered in previous analyses---
that plays a central role.
In our view, the identification and  resolution of this ambiguity 
clarifies the whole issue, as we now proceed to discuss. 

As in Ref.~\cite{Bramon:2005pz}, let's consider the two-photon entangled wave-function 
\begin{equation}
\label{psiy1y2def}
\Psi(y_1,y_2)=\int\int dk_1 dk_2\Psi(k_1,k_2) 
\frac{e^{i k_1 y_1}}{\sqrt{2\pi}}\frac{e^{ik_2y_2}}{\sqrt{2\pi}}\ , 
\end{equation}
where only the relevant, vertical components appear explicitly, 
$\hbar =1$, the integration limits are $\pm \infty$, and  
\begin{equation}
\label{psik1k2}
\Psi(k_1,k_2) =\frac{1}{\sqrt{\pi\sigma_+\sigma_-}} 
e^{-\frac{1}{4\sigma_+^2}\left(k_1+k_2\right)^2}
e^{-\frac{1}{4\sigma_-^2}\left(k_1-k_2\right)^2}\ .
\end{equation}
State (\ref{psik1k2}) can be Fourier-transformed into
\begin{equation}
\label{psiy1k2}
\Psi(y_1,k_2)=\sqrt{\frac{2}{\pi}\frac{\sigma_+\sigma_-}{\sigma_+^2 +\sigma_-^2}}
e^{- \frac{\sigma_+^2\sigma_-^2 y_1^2+k_2^2-i(\sigma_+^2-\sigma_-^2)y_1 k_2}
{\sigma_+^2+\sigma_-^2}},
\end{equation}
and
\begin{equation}
\label{psiy1y2}
\Psi(y_1,y_2)=\sqrt{\frac{\sigma_+ \sigma_-}{\pi}}
e^{-\frac{1}{4}\left[ \sigma_{+}^2 (y_1+y_2)^2+\sigma_{-}^2 (y_1-y_2)^2 \right]}\ . 
\end{equation}

Note that for the global system we have chosen a Gaussian wave packet  with
$\Delta(k_1+k_2)=\sigma_+$ and $\Delta\left[(y_1+y_2)/2 \right]=1/ 2\sigma_+$.
This allows for analytical computations and one has
$\Delta(k_1+k_2)\Delta\left[(y_1+y_2)/2\right]=1/2$,
which is the minimum value compatible with Heisenberg's principle.
In this sense, our state is the closest analog to Popper's  proposal,
which naively assumed a perfectly fixed and localized source.
It also coincides with the state discussed  by Qureshi \cite{Qureshi2,Qureshi3}
once $\sigma_+^2$ and $\sigma_-^2$ are substituted by $1/ \Omega_0^2$ and $4\sigma^2$.  
A Gaussian wave packet is also quite appropriate to discuss the  ``ghost'' diffraction
experiments \cite{Kim:1999td,Strekalov}, where an entangled  two-photon state is
obtained by spontaneous parametric down conversion (SPDC);
indeed, the pump laser which plays the role of the decaying system has an approximate
Gaussian profile.
Although the two SPD converted photons do not travel with opposite momenta 
($\vec{k_{1}}+\vec{k_{2}}=\vec{k}_{\rm pump}$), an ``unfolded'' version of the experimental scheme
(see Refs.~\cite{Kim:1999td,Strekalov}) is known to be equivalent to the schemes in Fig.~1.
On the other hand, the width of the pump beam is so large (a few mm) and its divergence so small (less than a mrad) that the SPD converted two-photon states in Refs.~\cite{Kim:1999td,Strekalov}
can be modelled by the state (\ref{psiy1y2def}) with
a large $\Delta \left[(y_1+y_2)/2 \right]$ and a small $\sigma_+\ll\sigma_-$. 

We thus assume that the state (\ref{psiy1y2def}) describes our system when
the left-moving photon reaches the screen in the setups of Fig.~1 \cite{tnotneed}.
Since we are going to restrict the discussion to jointly detected photon pairs and
this requires the passage of the left-photon through the slit,
previous analyses (including Popper's) consider that this effective passage through
the slit  amounts to performing a \textit{measurement} of the $y$-coordinate of the
left-moving photon with an accuracy $\Delta y_1\simeq a$. 
In our view, this is not exactly the case and generates important ambiguities. 
Strictly speaking, the effective passage through the left-slit amounts to perform a
\textit{state preparation} for the \textit{global} system one is going to work with.
The two photons remain entangled and none has a (pure) quantum state by itself until a measurement (including photon detection) is really performed.
But there are many possibilities for such a measurement and, somehow unexpectedly,
they lead to inequivalent situations. We concentrate on the two more interesting cases
of measurement: central \textit{versus} inclusive detections.

\paragraph{Central detection:}
One possibility consists in collecting coincident events with the left photon reaching 
the \textit{central} detector on the negative $x$-axis far left from the
slit.  This amounts to select left-photons with a measured $k_1=0$.
When one of these photons passes through the $2a$-wide slit, its $y$-coordinate is
established  with an accuracy $\Delta y_1\simeq a$.  
Then the indistinguishable photon paths covering the whole allowed range,  $-a \le y \le +a$,
reach the central detector in phase and the corresponding amplitudes add coherently thus
building the  zero-th order diffraction maximum.  
Once the left-moving photon is so detected, \textit{and only then},
its corresponding right-partner has a definite, pure quantum state and
QM can be used to make unambiguous and complete predictions for measurements
performed on it. 

In particular, for the vertical momentum of these photons
QM predicts a vanishing expectation value,  $\langle k_2\rangle=0$,
and a variance $(\Delta k_2)^2$ given by 
\begin{equation}
\label{deltak2def1}
(\Delta k_2)^2_a|_{CD}= 
\frac{\int_{-\infty}^{+\infty} dk_2 k_2^2 |\int_{-a}^{+a}dy_{1}\Psi(y_1;k_2)|^2}
        {\int_{-\infty}^{+\infty} dk_2 |\int_{-a}^{+a}dy_{1}\Psi(y_1;k_2)|^2}\ ,
\end{equation}
with integrals covering the opening of the left-slit and introducing the $a$-dependence
we are looking for.
For reasonably narrow slit-widths one can expand the Gaussian functions and retain the
first three terms to obtain
\begin{equation}
\label{deltak2v1}
(\Delta k_2)_a|_{CD}=\frac {1}{2} \sqrt{\sigma_+^2+\sigma_-^2}
\left[1-\frac{a^2}{12} \frac{(\sigma_+^2-\sigma_-^2)^2}{(\sigma_+^2+\sigma_-^2)}\right] \ . 
\end{equation}
Therefore, QM predicts an increase in the momentum spread of the right-moving photon
when narrowing the left-slit width,
which is precisely the same conclusion that Popper himself derived from standard QM.
Since there are no reasons to modify Popper's conclusions about the opposite behaviour
predicted by his own approach, we agree with Popper in that the test is \textit{crucial}.
At least, if the experiment is performed as just described.

And, as far as we know, present day experiments related to Popper's test are performed 
in this way. 
One is the experiment by Kim and Shih \cite{Kim:1999td},
which explicitly claims to be a realization of Popper's proposal.
The other is due to Strekalov \textit{et al.}~\cite{Strekalov} and makes no such a claim,
but has been recently interpreted by Qureshi \cite{Qureshi3} as a real Popper's test as well.
In both experiments, pairs of optical photons are jointly detected.
One member of each pair (corresponding to our left-moving photon) triggers a
\textit{fixed} detector centered on the optical axis, while the momentum of its partner
(our right-moving photon) is measured by a second detector scanning the region covered by
the right-array of detectors in Figs.~1a,b.
Our discussion should thus apply to these experimental setups and the meaning of their
results will be commented after considering a  second detection possibility.

\paragraph{Inclusive detection:}
Alternatively, the final left-photon detection could be performed not only by the central detector,
as before, but by the whole array of left-side detectors.
One thus collects a larger set of jointly detected events.
For each one of these events, a measurement of the vertical momenta of both members,
$k_1$ and $k_2$, has been performed.
Symmetry considerations and QM still predict the same vanishing $\langle k_2\rangle=0$
but the right-photon momentum spread $\Delta k_2$ turns out be 
different from the previous one, Eq.~(\ref{deltak2v1}).
It can be computed by noticing that one has to sum over all the $k_1$-values
measured for the photons passing through the left-slit and that this is equivalent to compute    
\begin{equation}
\label{deltak2def2}
(\Delta k_2)^2_a|_{ID}= 
\frac{\int_{-\infty}^{+\infty} dk_2 k_2^2 \int_{-a}^{+a}dy_{1}|\Psi(y_1;k_2)|^2}
        {\int_{-\infty}^{+\infty} dk_2 \int_{-a}^{+a}dy_{1}|\Psi(y_1;k_2)|^2}\ ,
\end{equation}
because the measurement one chooses to perform on the left-photon
(the precise vertical position $y_1$ when passing through the slit, Eq.~(\ref{deltak2def2}),
rather than its vertical momentum $k_1$ when reaching a detector)
cannot modify the (one side) probability distributions
(and thus the value for $\Delta k_2$) of the right-moving photons. 
Eqs.~(\ref{psiy1k2}) and (\ref{deltak2def2}) lead immediately to 
\begin{equation}
\label{deltak2v2}
(\Delta k_2)_a|_{ID}= \frac{1}{2}\sqrt{\sigma_+^2+\sigma_-^2}\ , 
\end{equation}
with {\it no} $a$-dependence and contrasting again with the predictions of Popper's
approach. The test is still {\it crucial}.  

We have thus identified an ambiguity in Popper's proposal which has to be solved
in order to obtain the correct QM prediction on the behaviour of $\Delta k_2$ when   
narrowing $a$. $\Delta k_2$ is predicted to increase for right-detections conditioned to 
the outcome $k_1 =0$ corresponding to a
`click' of the {\it central} detector on the left. But for coincidence detections with
some other (asymmetrically placed) left-detectors 
($k_1 \not= 0$), $\Delta k_2$ has to decrease and Popper's
test is no longer conclusive. The two opposite behaviours compensate and $\Delta k_2$ is
predicted to be $a$-independent for \textit{inclusive} detection over the whole left-array, 
{\it i.e.}, when one decides to ignore the existing information on the left-photon
momenta.    

An important question is to decide which scheme 
---with central or inclusive detection--- is closer to Popper's original proposal.
But there seems to be no clear answer for this question.
The main reason has  been anticipated by Qureshi \cite{Qureshi3}. 
The central point in Popper's argument is that the accuracy of the knowledge that
one can have  on the photon vertical coordinates is fixed
by the (left) slit-width, $\Delta y_1\simeq a$. 
As long as the set of collected  photons (on the right-side) are those whose other partner
passed through the (left) slit, they should show the effects that Popper was looking for.
But this is a common feature in both detection schemes and none can thus be favoured.  
One could argue that the schematic diagrams in Popper's writings show
symmetrical arrays of detectors on both sides and that their purpose,
apart from exhibiting diffraction by the physical slits in the setup of  Fig.~1a,
is also to collect all the photons crossing the slit in Fig.~1b.
This could be true, but the fact that in this case one has to ignore 
the information obtained on the left-photon momenta disfavours this interpretation of
Popper's argument,
an argument centered precisely on exploiting the available information. 
In our view, between the two considered options 
---a first one using all the information but not the full set of events \textit{versus}
a second one using all the events but not the full information---
none is clearly favoured.

Let's finally discuss the available experimental information coming from
Refs.~\cite{Kim:1999td,Strekalov} where, as previously mentioned,
photons passing through the slit are collected only by the central detector 
and has thus to be analyzed in terms of  Eq.~(\ref{deltak2v1}). 
Unfortunately, in the experimental realization of Popper's test by Kim and Shih \cite{Kim:1999td}
the width of the slit is kept fix, $2a = 0.16$ mm.
This precludes  any direct conclusion via Eq.~(\ref{deltak2v1}).
Note, however, that the slit width is much narrower than the incident beam and that
the use of such an extended source (a wide beam) is a way to respect 
Heisenberg's principle for the source and to circumvent the conventional 
criticisms previously mentioned.
The experimental findings in Ref.~\cite{Kim:1999td} have been clearly interpreted by Short \cite{Short}
but they are of marginal relevance for our present discussion. 
By contrast, the experiment by Strekalov \textit{et al.}~\cite{Strekalov}
is more interesting because the slit-width is modified (from 0.1 and 1.1 mm)
and this allows to discriminate between the $a$-dependence predicted by standard QM,
Eq.~(\ref{deltak2v1}), and the opposite behaviour attributed by Popper to his own approach. 
The data of Ref.~\cite{Strekalov} clearly show that $\Delta k_2$ increases when $a$ is decreased
and are compatible with the QM prediction in Eq.~(\ref{deltak2v1}).
Contrary to what Popper was ``inclined to predict'',
this experiment contradicts his own approach and vindicates conventional QM.

The same conclusion has recently been reached by Qureshi \cite{Qureshi3}
and was expected by many authors (see, \textit{e.g.}~\cite{Krips,Bramon:2005pz}).
But our present discussion shows that these claims were not well justified. 
Admittedly, they have turned out to be correct, but this is a consequence of the set of
photon pairs ---conditioned to central left-detection--- used in
Ref.~\cite{Strekalov}. For some other non-centered detectors, coincident events should 
show the opposite behaviour: $\Delta k_2$ increasing with $a$, as in Popper's
approach but without contradicting QM ---as previous analyses would
suggest.
Finally, if the set is enlarged to include all the events reaching whatever
detector behind the (left) slit, QM predicts that $\Delta k_2$ is $a$-independent, 
Eq.~(\ref{deltak2v2}), contradicting Popper's approach and making the test conclusive
again. There seems to be no data for this case and an experiment will be welcome.
After all, Popper's test and QM itself are really subtle,
as seen in the present paper and as brilliantly emphasized by Asher Peres \cite{Peres}.

We thank, V.~Ahufinger, K.~Eckert and  A.~J.~Short
for interesting comments.
This work is  partly supported by the Ram\'on y Cajal program (R.E.),
the Ministerio de Ciencia y Tecnolog\'{\i}a, BFM-2002-02588,
and the EU, HPRN-CT-2002-00311, EURIDICE network.

\end{document}